\documentclass[12pt]{article}

\usepackage{amssymb}
\usepackage{citesort}
\usepackage{amsmath,mathrsfs}
\usepackage{url}

\def\pa{\partial}
\def\fr{\frac}

\newtheorem{iden}{Identity}

\newcommand{\db}{de$\,$Broglie}
\newcommand{\Db}{De$\,$Broglie}

\newcommand{\dbb}{de$\,$Broglie-Bohm}
\newcommand{\Dbb}{De$\,$Broglie-Bohm}
\newcommand{\Endproof}{\hfill $\square$}
\newcommand{\ui}{\textrm{i}}

\begin{document}
\vspace*{1.0cm}
\noindent
{\bf
{\large
\begin{center}
\Dbb\ Guidance Equations for Arbitrary Hamiltonians
\end{center}
}
}

\vspace*{.5cm}
\begin{center}
Ward Struyve\\
Perimeter Institute for Theoretical Physics,\\
31 Caroline Street North, Waterloo, Ontario N2L 2Y5, Canada\\
and\\
Instituut voor Theoretische Fysica, K.U.Leuven\\
Celestijnenlaan 200D, B-3001 Leuven, Belgium.{\footnote{Present address. Currently Postdoctoral Fellow FWO.}}\\
E--mail: Ward.Struyve@fys.kuleuven.be
\end{center}

\begin{center}
Antony Valentini\\
Theoretical Physics Group, Blackett Laboratory, Imperial College London,\\
Prince Consort Road, London SW7 2AZ, United Kingdom.\\
E--mail: a.valentini@imperial.ac.uk
\end{center}

\begin{abstract}
\noindent
In a pilot-wave theory, an individual closed system is described by a wavefunction $\psi(q)$ and configuration $q$. The evolution of the wavefunction and configuration are respectively determined by the Schr\"odinger and guidance equations. The guidance equation states that the velocity field for the configuration is given by the quantum current divided by the density $|\psi(q)|^2$. We present the currents and associated guidance equations for any Hamiltonian given by a differential operator. These are derived directly from the Schr\"odinger equation, and also as Noether currents arising from a global phase symmetry associated with the wavefunction in configuration space.   
\end{abstract}

\renewcommand{\baselinestretch}{1.1}
\bibliographystyle{unsrt}

\section{Introduction}
In the pilot-wave theory of \db\ and Bohm \cite{debroglie28,bacciagaluppi06,bohm52a,bohm52b,bell87,holland93b,bohm93}, an individual closed system of nonrelativistic particles is described by a wavefunction $\psi({\bf x}_1,\dots,{\bf x}_N,t)$, which satisfies the nonrelativistic Schr\"odinger equation
\begin{equation}
\ui \hbar \partial_t \psi({\bf x}_1, \dots,{\bf x}_N,t) =  \left( -\sum^N_{k=1} \frac{\hbar^2 }{2m_k}\nabla^2_k + V({\bf x}_1, \dots,{\bf x}_N) \right)  \psi({\bf x}_1, \dots,{\bf x}_N,t),
\label{0.1}
\end{equation}
and by $N$ particle positions ${\bf x}_1,\dots,{\bf x}_N$, for which the possible trajectories are solutions of the guidance equations
\begin{equation}
\fr{d {\bf x}_k}{dt}= \frac{\hbar}{2\ui m_k|\psi|^2} \left( \psi^* {\boldsymbol {\nabla}}_k \psi -\psi{\boldsymbol {\nabla}}_k \psi^*   \right)   = \frac{1}{m_k} {\boldsymbol {\nabla}}_k S ,
\label{0.2}
\end{equation}
where $\psi=|\psi|\exp(\ui S/\hbar)$. Here, $\psi$ is regarded as an objective physical field in configuration space, guiding the motion of an individual system.

The `density' $|\psi|^2$ is preserved by the particle flow, in the sense that if, over an ensemble of systems with the same wavefunction $\psi$, the configurations have the `quantum equilibrium' distribution $|\psi({\bf x}_1, \dots,{\bf x}_N,t_0)|^2$ at time $t_0$, then they will have the distribution $|\psi({\bf x}_1, \dots,{\bf x}_N,t)|^2$ at any time $t$. This is a consequence of the continuity equation for $|\psi|^2$, 
\begin{equation}
\partial_t |\psi|^2  + \sum^N_{k=1}{\boldsymbol {\nabla}}_k \cdot \left(  \frac{{\boldsymbol {\nabla}}_k S}{m_k} |\psi|^2   \right) =0 ,
\label{0.3}
\end{equation} 
which itself follows from the Schr\"odinger equation.{\footnote{Note, however, that in principle the theory allows `nonequilibrium' ensemble distributions, that is, distributions that differ from $|\psi|^2$ \cite{valentini91a,valentini91b,valentini92,valentini062}.}}

\Db\ arrived at the guidance equation \eqref{0.2}, in the 1920s, by combining the variational principles of Maupertuis and Fermat. Bohm, on the other hand, who rediscovered \db's theory in the early 1950s, proposed an acceleration-based equation of motion for the particles (Newton's equation of motion with a `quantum potential'), instead of the velocity-based guidance equation. In Bohm's view, the guidance equation is regarded as a mere constraint on the initial momenta, a constraint that can in principle be relaxed \cite{bohm52a}. Whereas, in \db's view, the guidance equation is regarded as the fundamental law of motion. (For a full discussion, see ref.\ \cite{bacciagaluppi06}.)

However, \db's velocity field as given by the right-hand side of \eqref{0.2} cannot be correct in general. For while the resulting flow preserves $|\psi|^2$ for standard Hamiltonians of the form appearing in \eqref{0.1}, it will not preserve $|\psi|^2$ for Hamiltonians of an arbitrary form. For more general Hamiltonians, an acceptable guidance equation may be derived from the requirement that it preserves the density $|\psi|^2$ (a method that is in fact often used), since this property plays a key role in showing that in equilibrium the theory reproduces the standard quantum predictions. Given a continuity equation for a density $|\psi|^2$, one may define an associated `current'. The velocity field is then postulated to equal the current divided by the density.

In this paper we derive currents, from which the guidance equations can be constructed, for any Schr\"odinger equation such that the Hamiltonian is given by a differential operator. More explicitly, we consider Schr\"odinger equations{\footnote{Here and below we take $\hbar = 1$.}} 
\begin{equation}
\ui \partial_t  \psi(q,t) = {\widehat H}(q,\nabla,t) \psi(q,t),
\label{0.5}
\end{equation}
where $q=(q_1,\dots,q_N)\in {\mathbb{R}}^N$ and $\nabla$ is the $N$-dimensional gradient, and where ${\widehat H}(q,\nabla,t)$ is a differential operator, that is, a multivariate polynomial in the symbols $\pa_{q_1}, \dots, \pa_{q_N}$ with coefficients being complex-valued functions of $q$ and $t$. From \eqref{0.5}, we derive a continuity equation for the density $|\psi(q,t)|^2$ with a current $j^\psi(q,t)$, that is,
\begin{equation}
\partial_t |\psi(q,t)|^2 + \nabla \cdot j^\psi(q,t) = 0. 
\label{0.4}
\end{equation} 
A possible guidance equation is then given by 
\begin{equation}
\frac{dq}{dt} = \frac{j^\psi}{|\psi|^2}.
\label{0.6}
\end{equation}
Of course, one can always add a divergence-free vector field to $j^\psi(q,t)$ without losing the validity of \eqref{0.4}. However, it is standard to assume that the current vanish for $|q| \to + \infty$.

Note that, in the special case of one dimension ($N=1$), the requirement that the current vanishes at infinity uniquely determines the current to be
\begin{equation}
 j^\psi(q,t) = - \int^q_{-\infty} dq' \partial_t |\psi(q',t)|^2.  
\label{0.61}
\end{equation}
Other possible currents differ only by an additive $q$-independent term and hence do not vanish for $|q| \to + \infty$.

As we shall see, for a Hamiltonian that is a differential operator, we can always find a current $j^\psi(q)$ that is a local functional of the wavefunction $\psi(q)$ (that is, a $j^\psi(q)$ that can be written as a function of $q$, $\psi(q)$ and $q$-derivatives of $\psi(q)$ up to some finite order). In the uniquely-determined one-dimensional case ($N=1$), this implies that the expression on the right-hand side of \eqref{0.61} is a local functional. On the other hand, in the multi-dimensional case ($N \geqslant 2$) one can also find currents satisfying the continuity equation that are not local functionals. Here is an example. From the Schr\"odinger equation \eqref{0.5} it follows that
\begin{equation}
\partial_t |\psi(q,t)|^2 + I^\psi(q,t) =0 ,
\label{0.7}
\end{equation}
where
\begin{equation}
I^\psi(q,t) \equiv 2 \textrm{Re } \left(\ui\psi^*(q,t)  {\widehat H}(q,\partial_q,t) \psi(q,t) \right) .
\label{0.8}
\end{equation}
We can then write 
\begin{equation}
I^\psi(q,t) = \nabla^2 \left( \nabla^{-2}  I^{\psi}(q,t) \right) = \nabla \cdot \left(  \nabla \left( \nabla^{-2}  I^{\psi}({\bf x},t)\right) \right) ,
\label{0.9}
\end{equation}
where $\nabla^{-2}  I^{\psi}(q,t) = \int d^Nq' G_n(q-q')I^{\psi}(q',t) $ (assuming that the integral is well-defined), with 
\begin{equation}
G_N(q) = \left\{ \begin{array}{ll}
\frac{1}{2\pi} \log{|q|} , & N=2\\
-\frac{\Gamma(N/2- 1)}{4\pi^{N/2}|q|^{N-2}}, & N \geqslant 3
\end{array} \right.  
\label{0.91}
\end{equation}
a Green's function (a fundamental solution) for Laplace's equation in $N$ dimensions, that is, $\nabla^2 G_N(q) = \delta(q)$, see for example refs.\ \cite[pp.\ 49-53]{stackgold00} or \cite[pp.\ 21-25]{evans98}. The current 
\begin{equation}
j^\psi(q,t)=\nabla \left( \nabla^{-2}  I^{\psi}(q,t)\right)
\label{0.92}
\end{equation}
then satisfies the continuity equation. This method of obtaining a current was first suggested by Epstein \cite{epstein53} (for $N=3$). Although a local current is arguably preferable when the Hamiltonian is a differential operator, nonlocal currents of the type \eqref{0.92} could be useful for other Hamiltonians, such as the positive-energy Klein-Gordon Hamiltonian ${\sqrt{- \nabla^2  + m^2 }}$ (which is defined by transforming to Fourier space). 

Different choices of current lead to different guidance equations and hence to different pilot-wave models.{\footnote{Note, however, that merely postulating a velocity field by dividing a current $j^\psi$ by the density $|\psi|^2$ might not be sufficient to obtain a satisfactory pilot-wave model (that is, a pilot-wave model that reproduces the standard quantum predictions).}} Even requiring the current to be a local functional (insofar as the Hamiltonian admits that) does not resolve the ambiguity, as again the current is determined by the continuity equation only up to a divergence-free part. Deotto and Ghirardi \cite{deotto98} considered the latter ambiguity in detail, and showed that even imposing standard space-time symmetries still does not uniquely fix the current.  

Before considering the general case of a multi-dimensional configuration space, we shall first consider the one-dimensional case. This allows us to illustrate the techniques involved.

\section{One-dimensional configuration space}
Consider a Schr\"odinger equation
\begin{equation}
\ui \partial_t  \psi(q,t) = {\widehat H}(q,\partial_q,t) \psi(q,t) ,
\label{1}
\end{equation}
where $q \in {\mathbb R}$ and where  ${\widehat H}(q,\partial_q,t)$ is a differential operator, that is
\begin{equation}
{\widehat H}(q,\partial_q,t)\psi(q,t) = \sum_{n\geqslant 0} h_n(q,t) \partial^n_q\psi(q,t),
\label{2}
\end{equation}
where only a finite number of the complex-valued functions $h_n$ are non-zero. We assume that ${\widehat H}$ is Hermitian, that is, for wavefunctions $\psi_1$ and $\psi_2$,
\begin{equation}
\int dq \psi^*_1 {\widehat H} \psi_2 = \int dq ({\widehat H} \psi_1)^*  \psi_2 .
\label{2.01}
\end{equation}
From \eqref{2}, Hermiticity of ${\widehat H}$ implies
\begin{equation}
{\widehat H}(q,\partial_q,t)\psi(q,t) = \sum_{n \geqslant 0} (-\partial_q)^n \left( h^*_n(q,t)\psi(q,t) \right)
\label{2.1}
\end{equation}
(assuming that terms of the form $\partial^m_q(\psi_1 h_n)\partial^{n-m-1}_q \psi_2$ vanish at infinity). As shown in appendix \ref{appone}, it then follows that the Hamiltonian is Hermitian if and only if the coefficients $h_n$ satisfy
\begin{equation}
h_n = \sum_{m \geqslant n} (-1)^m  \binom{m}{n} \partial_q^{m-n}  h^*_m.
\label{3}
\end{equation}

From \eqref{1} it follows that
\begin{equation}
\partial_t |\psi(q,t)|^2 + I^\psi(q,t) =0 ,
\label{4}
\end{equation}
where
\begin{equation}
I^\psi(q,t) \equiv 2 \textrm{Re } \left(\ui\psi^*(q,t)  {\widehat H}(q,\partial_q,t) \psi(q,t) \right) .
\label{5}
\end{equation}
The goal is to write $I^\psi(q,t)$ in the form $\partial_q j^\psi(q,t)$.

In section~\ref{1dschrodinger} we derive the current $j^\psi$ by direct calculation. In section \ref{1dnoether} we derive it as a Noether current. We will make use of the following identity (the proof is given in appendix \ref{appone}):

\begin{iden}\label{iden1} For any two wave functions $\phi(q)$ and $\chi(q)$, and $n \geqslant 0$:{\footnote{Here and below it is understood that a sum is zero when the upper bound is smaller than the lower bound.}}
\begin{equation}
\phi \partial^n_q \chi - (-1)^n  \chi  \partial^n_q  \phi = \partial_q \left( \sum^{n-1}_{m=0} (-1)^{m} \partial^m_q \phi \partial^{n-m-1}_q \chi \right) .
\label{6}
\end{equation}

\end{iden}

\subsection{Deriving the current from the Schr\"odinger equation}\label{1dschrodinger}
We have 
\begin{align}
I^\psi &= 2 \textrm{Re } \left( \ui\psi^*   {\widehat H} \psi  \right)\nonumber\\
&= \ui\left( \psi^*   {\widehat H} \psi  -  \psi   \left({\widehat H}  \psi\right)^*\right)\nonumber\\
&= \ui \sum_{n \geqslant 1} \left( \psi^*  h_n  \partial^n_q  \psi   - (-1)^n \psi  \partial^n_q (h_n  \psi^*  ) \right),
\label{12}
\end{align}
where for ${\widehat H} \psi$ and $\left({\widehat H}  \psi\right)^*$ we have used respectively \eqref{2} and \eqref{2.1}. Making use of Identity~\ref{iden1} (equation \eqref{6}) for $\phi=\psi^* h_n$ and $\chi = \psi$, we can write $I^\psi  = \partial_q j^\psi $, with 
\begin{equation}
j^\psi  = \ui \sum_{n \geqslant 1} \sum^{n-1}_{m=0} (-1)^{m} \partial^m_q(\psi^*  h_n ) \partial^{n-m-1}_q \psi.
\label{13}
\end{equation}

We can also write the current in the form
\begin{equation}
j^\psi  = \sum_{n,m \geqslant 0} J_{nm}  \partial^{n}_q \psi \partial^m_q \psi^*,
\label{14}
\end{equation}
where the coefficients 
\begin{equation}
J_{nm} =  \ui\sum_{r \geqslant n+m+1} (-1)^{r+n+1} \binom{r-n-1}{m} \partial^{r-n-m-1}_q h_r ,\quad n,m \geqslant 0.
\label{15}
\end{equation}
are completely determined by the Hamiltonian (the proof of \eqref{15} is given in appendix \ref{appone}).

The above current $j^\psi$ is real. One way to see this is by noting that the current on the right-hand side of \eqref{13} (which vanishes for $|q| \to +\infty$) equals that on the right-hand side of \eqref{0.61}, which is obviously real. For the case of a multi-dimensional configuration space, we do not know of any simple argument that shows the reality of the obtained current. Instead, in appendix \ref{appmulti} we prove the reality by direct calculation, using the Hermiticity of the Hamiltonian. Since our discussion of the one-dimensional case serves as preparation for the multi-dimensional case, in appendix \ref{appone} we give a similar reality proof for the one-dimensional case.

\subsection{Deriving the current as a Noether current}\label{1dnoether}
In the previous section we derived the current $j^\psi$ by direct calculation from the Schr\"odinger equation. In this section the same current is derived as a Noether current. 

As is well known, Noether's theorem states that, for theories that can be derived from an action principle, each continuous symmetry implies a conserved quantity. Here the conserved quantity is quantum probability, and the conservation law is expressed in terms of the continuity equation. The Schr\"odinger equation \eqref{1} can be derived from an action, and the invariance of the action under global phase transformations $\psi \to e^{i\alpha}\psi$ implies the conservation of quantum probability.

Specifically, the Schr\"odinger equation \eqref{1} can be derived from an action principle with Lagrangian
\begin{equation}
L[\psi,{\dot{\psi}},\psi^*,{\dot{\psi}}^*,t]=\int dq \left(\frac{\ui}{2}(\psi^*(q,t){\dot{\psi}}(q,t) - {\dot{\psi}}^*(q,t) \psi(q,t)) - \psi^*(q,t) {\widehat H}(q,\partial_q,t) \psi(q,t) \right) 
\label{16}
\end{equation}
(a functional of $\psi,{\dot{\psi}},\psi^*,{\dot{\psi}}^*$), where ${\dot{\psi}} = \partial_t \psi$. The Euler-Lagrange equations are given by
\begin{equation}
\partial_t \frac{\delta L}{\delta {\dot{\psi}}} - \frac{\delta L}{\delta \psi} =0,\quad  \partial_t \frac{\delta L}{\delta {\dot{\psi}}^*} - \frac{\delta L}{\delta \psi^*} =0
\label{17}
\end{equation}
and just yield the Schr\"odinger equation and its complex conjugate.

The corresponding Lagrangian density may be taken as the function (using the Hermiticity of ${\widehat H}$)
\begin{multline}
{\mathcal L} (q, {\dot{\psi}}(q,t), {\dot{\psi}}^*(q,t),\psi(q,t),\psi^*(q,t),\dots,\partial^n_q \psi(q,t),\partial^n_q \psi^*(q,t),\dots,t) \\
 = \frac{1}{2} \left( \ui \psi^*(q,t){\dot{\psi}}(q,t) - \psi^* (q,t){\widehat H}(q,\partial_q,t) \psi(q,t)  + {\textrm{c.c.}}  \right)
\label{18}
\end{multline}
and in terms of ${\mathcal L}$ the Euler-Lagrange equations \eqref{17} read
\begin{equation}
\partial_t \frac{\partial {\mathcal L}}{\partial {\dot{\psi}}} - \sum_{n \geqslant 0} (-1)^n \partial^n_q  \frac{\partial {\mathcal L}}{\partial \partial^n_q  \psi}=0,\quad  \partial_t \frac{\partial {\mathcal L}}{\partial {\dot{\psi}}^*} - \sum_{n \geqslant 0} (-1)^n \partial^n_q  \frac{\partial {\mathcal L}}{\partial \partial^n_q  \psi^*}=0.
\label{19}
\end{equation}

Since the field-theoretical Noether theorem is usually discussed only for Lagrangian densities that depend on up to first-order derivatives of the field, we repeat the analysis in full. The generalization to Lagrangian densities that depend on higher-order derivatives is in fact straightforward. 

If ${\mathcal L}$ is invariant under an infinitesimal symmetry transformation $\psi \to \psi + \delta \psi$, $\psi^* \to \psi^* + \delta \psi^*$, we have $\delta {\mathcal L}=0$. We can write $\delta {\mathcal L}$ as
\begin{eqnarray}
\delta {\mathcal L} &=& \frac{\partial {\mathcal L}}{\partial {\dot{\psi}}}  \delta {\dot{\psi}} + \sum_{n \geqslant 0}   \frac{\partial {\mathcal L}}{\partial \partial^n_q  \psi} \delta (\partial^n_q\psi) + {\textrm{c.c.}} \nonumber\\
&=& \partial_t \left( \frac{\partial {\mathcal L}}{\partial {\dot{\psi}}}  \delta \psi \right) -  \partial_t \frac{\partial {\mathcal L}}{\partial {\dot{\psi}}}  \delta \psi + \sum_{n \geqslant 0}   \frac{\partial {\mathcal L}}{\partial \partial^n_q  \psi}  \partial^n_q(\delta\psi) + {\textrm{c.c.}}  \nonumber\\
&=&  \partial_t \left( \frac{\partial {\mathcal L}}{\partial {\dot{\psi}}}  \delta \psi \right)  + \sum_{n \geqslant 0}  \left(   \frac{\partial {\mathcal L}}{\partial \partial^n_q  \psi} \partial^n_q  (\delta\psi) - (-1)^n \partial^n_q \left( \frac{\partial {\mathcal L}}{\partial \partial^n_q  \psi} \right) \delta\psi    \right)  + {\textrm{c.c.}}, 
\label{20}
\end{eqnarray}
where in the last line we have used the Euler-Lagrange equations. Using Identity~\ref{iden1} (equation \eqref{6}) in the second term on the right hand side, with $\phi = \partial {\mathcal L}/ \partial  \partial^n_q  \psi$ and $\chi = \delta\psi$, we obtain
\begin{equation}
0= \delta {\mathcal L} =  \partial_t \left( \frac{\partial {\mathcal L}}{\partial {\dot{\psi}}}  \delta \psi \right)  + \partial_q \left( \sum_{n \geqslant 1} \sum^{n-1}_{m=0} (-1)^{m} \partial^m_q \left( \frac{\partial {\mathcal L}}{\partial \partial^n_q  \psi} \right) \partial^{n-m-1}_q (\delta\psi) \right) + {\textrm{c.c.}} 
\label{21}
\end{equation}
and hence we obtain a conserved `2-current', with density 
\begin{equation}
\rho^\psi = \frac{\partial {\mathcal L}}{\partial {\dot{\psi}}}  \delta \psi + {\textrm{c.c.}} 
\label{22}
\end{equation}
and `spatial' current
\begin{equation}
j^\psi = \sum_{n \geqslant 1} \sum^{n-1}_{m=0} (-1)^{m} \partial^m_q \left( \frac{\partial {\mathcal L}}{\partial \partial^n_q  \psi} \right) \partial^{n-m-1}_q (\delta\psi)  + {\textrm{c.c.}} 
\label{23}
\end{equation}
(where $(\rho^\psi,j^\psi)$ is determined up to an additive term with vanishing 2-divergence).

The current associated with the global phase symmetry $\psi \to \psi - \ui \varepsilon \psi$, $\psi^* \to \psi^* + \ui \varepsilon \psi^*$ is then given by
\begin{eqnarray}
\rho^\psi &=& \varepsilon |\psi|^2 , \label{24} \\
j^\psi &=& 2 {\textrm{Re }} \left[  -\ui \varepsilon\sum_{n \geqslant 1} \sum^{n-1}_{m=0} (-1)^{m} \partial^m_q \left( \frac{\partial {\mathcal L}}{\partial \partial^n_q  \psi} \right) \partial^{n-m-1}_q \psi    \right].
\label{25}
\end{eqnarray}
Since
\begin{equation}
\frac{\partial {\mathcal L}}{\partial \partial^n_q  \psi} =  -\frac{1}{2} \psi^* \frac{\partial  (  {\widehat H} \psi ) }{\partial \partial^n_q  \psi} , 
\label{26}
\end{equation}
for $n \geqslant 1$, the current \eqref{25} can also be written as
\begin{equation}
j^\psi =  {\textrm{Re }} \left[  \ui \varepsilon\sum_{n \geqslant 1} \sum^{n-1}_{m=0} (-1)^{m} \partial^m_q \left(  \psi^*\frac{\partial  (  {\widehat H} \psi ) }{\partial \partial^n_q  \psi}    \right) \partial^{n-m-1}_q \psi    \right].
\label{27}
\end{equation}
Using $\partial  (  {\widehat H} \psi )/ \partial \partial^n_q  \psi = h_n$, the expression in square brackets agrees with the right-hand side of \eqref{13}, up to the constant factor $\varepsilon$. It therefore also follows that taking the real part in equations \eqref{25} and \eqref{27} is redundant, as the argument of Re in those expressions is already real.

%%%%%%%%%%%%%%%%%%%%%%%%%%%%%%%%%%%%%%%%%%%%%%%%%%%%%%%%%%%%%%%%%%%%%%%%%%%%%%%%%%%%%%%%%%%%%%%%%%%55
%%%%%%%%%%%%%%%%%%%%%%%%%%%%%%%%%%%%%%%%%%%%%%%%%%%%%%%%%%%%%%%%%%%%%%%%%%%%%%%%%%%%%%%%%%%%%%%%%%%55
%%%%%%%%%%%%%%%%%%%%%%%%%%%%%%%%%%%%%%%%%%%%%%%%%%%%%%%%%%%%%%%%%%%%%%%%%%%%%%%%%%%%%%%%%%%%%%%%%%%%55

\section{$N$-dimensional configuration space}
We now turn to the case of an $N$-dimensional configuration space, with points $q=(q_1,\dots,q_N)\in {\mathbb R}^N$. We shall use multi-indices, which are elements of ${\mathbb N}^N_0$, that is, we use an index $n=(n_1,\dots,n_N)\in {\mathbb N}^N_0$. We define $e_i \in {\mathbb N}^N_0$, for $i=1,\dots,N$, as $(e_i)_j=\delta_{ij}$ ($j=1,\dots,N$). We also define $0 \in {\mathbb N}^N_0$ as the element $0=(0,\dots,0)$. We further assume the following standard definitions:
\begin{align}
& n \pm n' = (n_1 \pm n'_1,\dots,n_N \pm n'_N), \nonumber\\
& n\leqslant n' \quad {\textrm{if}} \quad  n_i \leqslant n'_i,\ i =1,\dots,N, \nonumber\\
& n < n' \quad {\textrm{if}} \quad  n\leqslant n' {\textrm{ and }} n \neq n', \nonumber\\
& |n| = \sum^N_{i=1} n_i, \nonumber\\
& n! = \prod^N_{i=1} (n_i!), \nonumber\\
& D^n = \pa^{n_1}_{q_1} \dots \pa^{n_N}_{q_N},\nonumber\\
& \binom{n}{n'} = \binom{n_1}{n'_1} \dots \binom{n_N}{n'_N},
\label{50}
\end{align}
where $n,n' \in  {\mathbb N}^N_0$.

Consider a Schr\"odinger equation
\begin{equation}
\ui \partial_t  \psi(q,t) = {\widehat H}(q,D,t) \psi(q,t) ,
\label{51}
\end{equation}
where ${\widehat H}(q,D,t)$ is a differential operator, that is 
\begin{equation}
{\widehat H}(q,D,t)\psi(q,t) = \sum_{n \geqslant 0} h_n(q,t) D^n\psi(q,t),
\label{52}
\end{equation}
where only a finite number of the complex-valued functions $h_n$ are non-zero. From \eqref{52}, Hermiticity of ${\widehat H}$ now implies
\begin{equation}
{\widehat H}(q,D,t)\psi(q,t) =\sum_{n \geqslant 0} (-1)^{|n|} D^n \left( h^*_n(q,t)\psi(q,t) \right)
\label{52.1}
\end{equation}
(assuming that terms of the form $D^m(\psi_1 h_n)D^{n-m-e_i} \psi_2$, for wavefunctions $\psi_1$ and $\psi_2$, vanish at infinity). As shown in appendix \ref{appmulti}, it then follows that the Hamiltonian is Hermitian if and only if the coefficients $h_n$ satisfy
\begin{equation}
h_n = \sum_{m \geqslant n} (-1)^{|m|}  \binom{m}{n} D^{m-n}  h^*_m.
\label{53}
\end{equation}

From \eqref{51} it follows that
\begin{equation}
\partial_t |\psi(q,t)|^2 + I^\psi(q,t) =0 ,
\label{54}
\end{equation}
where
\begin{equation}
I^\psi(q,t) \equiv 2 \textrm{Re } \left(\ui\psi^*(q,t)  {\widehat H}(q,D,t) \psi(q,t) \right) .
\label{55}
\end{equation}
The goal is to write $I^\psi(q,t)$ in the form $\sum^N_{i=1}  D^{e_i}  j^\psi_i(q,t)$.

In section~\ref{ndschrodinger} we derive a current $j^\psi(q)$ by direct calculation. In section \ref{ndnoether} we derive it as a Noether current. We will make use of the following identity (the proof is given in appendix \ref{appmulti}):
\begin{iden}\label{iden2} For any two wave functions $\phi(q)$ and $\chi(q)$, and $n \geqslant 0$:
\begin{multline}
\phi D^n \chi - (-1)^{|n|}    \chi  D^n  \phi\\
 =  \sum^N_{i=1} D^{e_i} \left( \sum_{0 \leqslant m \leqslant n - e_i} (-1)^{|m|} \frac{n!}{|n|!}  \frac{|m|!}{m!} \frac{|n-m-e_i|!}{(n-m-e_i)!}  D^m \phi D^{n - m - e_i} \chi \right).
\label{56}
\end{multline}
\end{iden}

\subsection{Deriving the current from the Schr\"odinger equation}\label{ndschrodinger}
We have
\begin{align}
I^\psi &= 2 \textrm{Re } \left( \ui\psi^*   {\widehat H} \psi  \right)\nonumber\\
&= \ui\left( \psi^*   {\widehat H} \psi  -  \psi   \left({\widehat H}  \psi\right)^*\right)\nonumber\\
&= \ui \sum_{n > 0}  \left( \psi^*  h_n  D^n  \psi   - (-1)^{|n|} \psi  D^n (h_n  \psi^*  ) \right),
\label{57}
\end{align}
where for ${\widehat H} \psi$ and $\left({\widehat H}  \psi\right)^*$ we have used respectively \eqref{52} and \eqref{52.1}. Making use of Identity~\ref{iden2} (equation \eqref{56}) for $\phi=\psi^* h_n$ and $\chi = \psi$, we can write $I^\psi  = \sum^N_{i=1} D^{e_i} j^\psi_i $, with
\begin{equation}
j^\psi_i  = \ui \sum_{n \geqslant e_i}    \sum_{0 \leqslant m \leqslant n - e_i} (-1)^{|m|} \frac{n!}{|n|!}  \frac{|m|!}{m!} \frac{|n-m-e_i|!}{(n-m-e_i)!}  D^m (\psi^* h_n)  D^{n - m - e_i} \psi. 
\label{58}
\end{equation}
(up to a divergence-free term). In appendix \ref{appmulti} we prove that this current is real.

We can also write the current in the form
\begin{equation}
j^\psi_i  = \sum_{n,m \geqslant 0} J_{i,nm}  D^n \psi D^m \psi^*,
\label{59}
\end{equation}
where the coefficients 
\begin{equation}
J_{i,nm} = \ui \sum_{r \geqslant n+m+e_i}  (-1)^{|r+n|+1}  \frac{r!}{|r|!} \frac{|r-n-e_i|!}{(r-n-e_i)!} \frac{|n|!}{n!} \binom{r-n-e_i}{m} D^{r-n-m-e_i} h_r ,
\label{60}
\end{equation}
with $n,m \geqslant 0$, are completely determined by the Hamiltonian (the proof of \eqref{60} is given in appendix \ref{appmulti}).

\subsection{Deriving the current as a Noether current}\label{ndnoether}
The current can also be derived as a Noether current associated with the global phase symmetry of the Lagrangian
\begin{equation}
L[\psi,{\dot{\psi}},\psi^*,{\dot{\psi}}^*,t]=\int d^N q \left(\frac{\ui}{2}(\psi^*{\dot{\psi}} - {\dot{\psi}}^* \psi) - \psi^* {\widehat H} \psi \right) ,
\label{61}
\end{equation}
where ${\dot{\psi}} = \partial_t \psi$. The Euler-Lagrange equations are given by
\begin{equation}
\partial_t \frac{\delta L}{\delta {\dot{\psi}}} - \frac{\delta L}{\delta \psi} =0,\quad  \partial_t \frac{\delta L}{\delta {\dot{\psi}}^*} - \frac{\delta L}{\delta \psi^*} =0,
\label{62}
\end{equation}
which yield the Schr\"odinger equation \eqref{51} and its complex conjugate.

The corresponding Lagrangian density may be taken as the function 
\begin{equation}
{\mathcal L} = \frac{1}{2} \left( \ui(\psi^*{\dot{\psi}} - {\dot{\psi}}^* \psi) - \psi^* {\widehat H} \psi  - \left({\widehat H} \psi\right)^* \psi\right).
\end{equation}
In terms of ${\mathcal L}$, the Euler-Lagrange equations \eqref{62} read
\begin{equation}
\partial_t \frac{\partial {\mathcal L}}{\partial {\dot{\psi}}} - \sum_{n \geqslant 0} (-1)^{|n|} D^n  \frac{\partial {\mathcal L}}{\partial D^n  \psi}=0,\quad  \partial_t \frac{\partial {\mathcal L}}{\partial {\dot{\psi}}^*} - \sum_{n \geqslant 0} (-1)^{|n|} D^n  \frac{\partial {\mathcal L}}{\partial D^n  \psi^*}=0.
\label{64}
\end{equation}

If ${\mathcal L}$ is invariant under an infinitesimal symmetry transformation $\psi \to \psi + \delta \psi$, $\psi^* \to \psi^* + \delta \psi^*$, we have $\delta {\mathcal L}=0$. We can write $\delta {\mathcal L}$ as
\begin{eqnarray}
\delta {\mathcal L} &=& \frac{\partial {\mathcal L}}{\partial {\dot{\psi}}}  \delta {\dot{\psi}} + \sum_{n \geqslant 0}   \frac{\partial {\mathcal L}}{\partial D^n  \psi} \delta (D^n\psi) + {\textrm{c.c.}} \nonumber\\
&=& \partial_t \left( \frac{\partial {\mathcal L}}{\partial {\dot{\psi}}}  \delta \psi \right) -  \partial_t \frac{\partial {\mathcal L}}{\partial {\dot{\psi}}}  \delta \psi + \sum_{n \geqslant 0}   \frac{\partial {\mathcal L}}{\partial D^n  \psi}  D^n(\delta\psi) + {\textrm{c.c.}}  \nonumber\\
&=&  \partial_t \left( \frac{\partial {\mathcal L}}{\partial {\dot{\psi}}}  \delta \psi \right)  + \sum_{n \geqslant 0}  \left(   \frac{\partial {\mathcal L}}{\partial D^n  \psi} D^n  (\delta\psi) - (-1)^{|n|} D^n \left( \frac{\partial {\mathcal L}}{\partial D^n  \psi} \right) \delta\psi    \right)  + {\textrm{c.c.}} ,
\label{65}
\end{eqnarray}
where in the last line we have used the Euler-Lagrange equations. Using Identity~\ref{iden2} (equation \eqref{56}) in the second term on the right hand side, with $\phi = \partial {\mathcal L}/ \partial  D^n  \psi$ and $\chi = \delta\psi$, we obtain
\begin{multline}
0= \delta {\mathcal L} =  \partial_t \left( \frac{\partial {\mathcal L}}{\partial {\dot{\psi}}}  \delta \psi \right) +  \sum^N_{i=1} D^{e_i} \Bigg( \sum_{n \geqslant e_i} \sum_{0 \leqslant m \leqslant n - e_i} (-1)^{|m|} \\
\times \frac{n!}{|n|!}  \frac{|m|!}{m!} \frac{|n-m-e_i|!}{(n-m-e_i)!}  D^m \left( \frac{\partial {\mathcal L}}{\partial D^n  \psi}  \right) D^{n - m - e_i} \delta\psi  \Bigg) + {\textrm{c.c.}} 
\label{66}
\end{multline}
and hence we obtain a conserved `$(N+1)$-current' (up to an arbitrary additive term with vanishing $(N+1)$-divergence), with density 
\begin{equation}
\rho^\psi = \frac{\partial {\mathcal L}}{\partial {\dot{\psi}}}  \delta \psi + {\textrm{c.c.}} 
\label{67}
\end{equation}
and `spatial' currents
\begin{equation}
j^\psi_i  =  \sum_{n \geqslant e_i}     \sum_{0 \leqslant m \leqslant n - e_i} (-1)^{|m| } \frac{n!}{|n|!} \frac{|m|!}{m!} \frac{|n-m-e_i|!}{(n-m-e_i)!}  D^m \left( \frac{\partial {\mathcal L}}{\partial D^n  \psi}  \right) D^{n - m - e_i} \delta\psi + {\textrm{c.c.}} .
\label{68}
\end{equation}

The current associated with the global phase symmetry $\psi \to \psi - \ui \varepsilon \psi$, $\psi^* \to \psi^* + \ui \varepsilon \psi^*$ is then given by
\begin{align}
\rho^\psi &= \varepsilon |\psi|^2 , \label{n0.9120} \\
j^\psi_i &= 2 {\textrm{Re }} \Bigg[  -\ui \varepsilon\sum_{n \geqslant e_i}   \sum_{0 \leqslant m \leqslant n - e_i} (-1)^{|m| } \frac{n!}{|n|!}  \frac{|m|!}{m!} \frac{|n-m-e_i|!}{(n-m-e_i)!}  D^m \left( \frac{\partial {\mathcal L}}{\partial D^n  \psi}  \right) D^{n - m - e_i} \psi    \Bigg]. 
\label{69}
\end{align}
Since
\begin{equation}
\frac{\partial {\mathcal L}}{\partial D^n  \psi} = -\frac{1}{2} \psi^* \frac{\partial  (  {\widehat H} \psi ) }{\partial D^n  \psi} , 
\label{70}
\end{equation}
for $n>0$, the current \eqref{69} can also be written as
\begin{multline}
j^\psi_i = {\textrm{Re }} \Bigg[  \ui \varepsilon\sum_{n \geqslant e_i} \sum_{0 \leqslant m \leqslant n - e_i} (-1)^{|m| }  \\
\times  \frac{n!}{|n|!}  \frac{|m|!}{m!} \frac{|n-m-e_i|!}{(n-m-e_i)!}  D^m \left(       \psi^*\frac{\partial  (  {\widehat H} \psi ) }{\partial D^n  \psi}    \right) D^{n - m - e_i} \psi  \Bigg].
\label{71}
\end{multline}
Using $\partial  (  {\widehat H} \psi )/ \partial D^n  \psi = h_n$, the expression in square brackets agrees with the right-hand side of \eqref{58}, up to the constant factor $\varepsilon$. It therefore also follows that taking the real part in equations \eqref{69} and \eqref{71} is again redundant as the argument of Re in those expressions is already real.

\section{Comparison with related work}
The construction of quantum probability currents and associated guidance equations, for certain classes of operators, has been considered before by a number of authors, in particular Brown and Hiley \cite{brown00,brown04}, D\"urr {\em et al.}\ \cite{durr031,durr04}, and Gambetta and Wiseman \cite{gambetta04}.{\footnote{Stone \cite{stone94} considers Hamiltonians of the form \eqref{2} where the $h_n$ are constant. Instead of writing the guidance equation in terms of derivatives of $\psi$ and $\psi^*$, Stone attempts to write it in terms of the real and imaginary parts of $(\ui\pa_q)^n\psi/\psi$. However, Stone's expression for the guidance equation appears to be incorrect. While it is rather straightforward to rewrite the guidance equation in terms of Stone's variables, this does not seem particularly useful. Therefore we do not present it here.}}

\subsection{Brown and Hiley}
Brown and Hiley consider one-dimensional Hamiltonians of the form
\begin{equation}
{\widehat H} = \sum_{m,n} h_{mn} {\widehat q}^m {\widehat p}^n,
\label{80}
\end{equation}
where the $h_{mn}$ are constant and ${\widehat p}$ is the momentum operator conjugate to ${\widehat q}$. In the configuration representation, where $\langle q| {\widehat q} | q' \rangle= q \delta(q-q')$, $\langle q| {\widehat p} | q' \rangle= -\ui \partial_q\delta(q-q')$, the Hamiltonian operator is given by 
\begin{equation}
\langle q| {\widehat H} | q' \rangle = {\widehat H}(q,\pa_q)\delta(q-q')= \sum_{m,n} h_{mn} q^m (-\ui\pa_q)^n \delta(q-q'),\label{80.1}
\end{equation}
 which corresponds to a Hamiltonian of the form \eqref{2} with $h_n = \sum_{m} h_{mn} q^m (-\ui)^n$.   

They find a current that can be written in the compact form 
\begin{equation}
j^\psi(q,t) = \langle q| \pa_{{\widehat p}}({\widehat  \rho}(t) {\widehat H}) | q \rangle  ,
\label{81}
\end{equation}
where ${\widehat  \rho}(t)= |\psi(t)\rangle \langle \psi(t) |$ is the density operator, and $\pa_{{\widehat p}}$ is the symbolic differential operator of Born and Jordan \cite{brown00,brown04,born25}, which acts on ${\widehat  \rho} {\widehat q}^m {\widehat p}^n$ as

\begin{equation}
\pa_{{\widehat p}} ({\widehat  \rho} {\widehat q}^m {\widehat p}^n) = \sum^n_{k=1}{\widehat p}^{n-k}{\widehat  \rho} {\widehat q}^m {\widehat p}^{k-1} ,
\label{82}
\end{equation}
and whose action is extended to ${\widehat  \rho}{\widehat H}$ by linearity. It can easily be shown that, in its domain of validity, this current coincides with that given in \eqref{13}.

It is straightforward to verify that Brown and Hiley's expression for the current also applies to Hamiltonians of the form  
\begin{equation}
{\widehat H} = \sum_{n} {\widehat g}_n(q,t) {\widehat p}^n,
\label{83}
\end{equation}
where $\langle q| {\widehat g}_n(q,t) | q' \rangle = g_n(q,t)\delta(q-q')$, and which correspond to Hamiltonians ${\widehat H}(q,\pa_q,t)$ of the form given in \eqref{2} with $h_n = g_n (-\ui)^n$. (The Born-Jordan derivative acts on ${\widehat  \rho} {\widehat g}_n(q,t) {\widehat p}^n$ in the same way as on ${\widehat  \rho} {\widehat q}^m {\widehat p}^n$, replacing ${\widehat q}^m$ by ${\widehat g}_n(q,t)$ in \eqref{82}.)

While we have not investigated this further, a similar form can probably be obtained in the case of a multi-dimensional configuration space.

\subsection{D\"urr {\em et al.}, Gambetta and Wiseman}
D\"urr {\em et al.}\ \cite{durr031,durr04} and, independently, Gambetta and Wiseman \cite{gambetta04} consider multi-dimensional Hamiltonians of up to second order in the momenta, that is Hamiltonians of the form (we follow a notation similar to that of Gambetta and Wiseman)
\begin{equation}
{\widehat H} = \sum_{i,j} {\widehat a}_{ij} {\widehat p}_i {\widehat p}_j + \sum_{i} {\widehat b}_{i} {\widehat p}_i + {\widehat c} ,
\label{84}
\end{equation}
where ${\widehat a}_{ij},{\widehat b}_{i},{\widehat c}$ are arbitrary time-dependent functions of the ${\widehat q}_i$. In the configuration representation, these correspond to Hamiltonians of the form \eqref{52} with the $h_n$ zero for $|n| > 2$.  

They find a current that can be written as
%{\footnote{Actually, Gambetta and Wiseman do not consider Hamiltonians with crossterms of the momentum operators. However, their result is straightforwardly generalized to the case where such crossterms are allowed.}} 
\begin{equation}
j^\psi_i(q,t) = {\textrm{Re}} \left( \langle \psi(t) | q \rangle \langle q | {\widehat v}_i(t) |\psi(t) \rangle \right) = {\textrm{Re}}  \langle q | {\widehat v}_i(t) {\widehat  \rho}(t)| q \rangle  ,
\label{85}
\end{equation}
where ${\widehat v}_k = \ui [{\widehat H}, {\widehat q}_k]$ is the velocity operator. This expression for the current is not valid for Hamiltonians containing higher-order terms in the momentum operators, corresponding to some non-zero $h_n$ for $|n| > 2$. However, in its domain of validity, it can be easily be shown to coincide with the current given in \eqref{58}.

%%%%%%%%%%%%%%%%%%%%%%%%%%%%%%%%%%%%%%%%%%%%%%%%%%%%%%%%%%%%%%%%%%%%%%%%%%%%%%%%%%%%%%%%%%%%%%%%%%%%%%%%%%%%%%%%%%5
%%%%%%%%%%%%%%%%%%%%%%%%%%%%%%%%%%%%%%%%%%%%%%%%%%%%%%%%%%%%%%%%%%%%%%%%%%%%%%%%%%%%%%%%%%%%%%%%%%%%%%%%%%%%%%%%%%5
%%%%%%%%%%%%%%%%%%%%%%%%%%%%%%%%%%%%%%%%%%%%%%%%%%%%%%%%%%%%%%%%%%%%%%%%%%%%%%%%%%%%%%%%%%%%%%%%%%%%%%%%%%%%%%%%%%5

\section{Conclusion}
We have presented \dbb\ guidance equations for any Schr\"odinger equation such that the Hamiltonian is given by a differential operator. These results may be applied to the development of possible pilot-wave interpretations for quantum theories with a Schr\"odinger time evolution. While we have considered only finite-dimensional configuration spaces, the equations carry over --- at least formally --- to field-configuration spaces. Thus the work can also be applied to the development of pilot-wave interpretations for quantum field theories, in terms of field ontologies.

\section{Acknowledgements}
W.S.'s research at Perimeter Institute for Theoretical Physics is supported by the Government of Canada through Industry Canada and by the Province of Ontario through the Ministry of Research \& Innovation. W.S.\ currently acknowledges the support of the FWO-Flanders. A.V.'s research is supported by grant RFP1-06-13A from The Foundational Questions Institute (\url{fqxi.org}). A.V.\ is grateful to Jonathan Halliwell for hospitality at Imperial College London. We thank a referee for helpful suggestions.

%%%%%%%%%%%%%%%%%%%%%%%%%%%%%%%%%%%%%%%%%%%%%%%%%%%%%%%%%%%%%%%%%%%%%%%%%%%%%%%%%%%%%%%%%%%%%%%%%%%%%%%%%%%%%%%%%%5
%%%%%%%%%%%%%%%%%%%%%%%%%%%%%%%%%%%%%%%%%%%%%%%%%%%%%%%%%%%%%%%%%%%%%%%%%%%%%%%%%%%%%%%%%%%%%%%%%%%%%%%%%%%%%%%%%%5
%%%%%%%%%%%%%%%%%%%%%%%%%%%%%%%%%%%%%%%%%%%%%%%%%%%%%%%%%%%%%%%%%%%%%%%%%%%%%%%%%%%%%%%%%%%%%%%%%%%%%%%%%%%%%%%%%%5

\appendix
\section{Proofs for a one-dimensional configuration space}\label{appone}

\noindent {\it Proof of the Hermiticity condition \eqref{3}.}
From \eqref{2.1}, and using Leibniz' rule, we have
\begin{equation}
{\widehat H}(q,\partial_q,t)\psi(q,t) = \sum_{n \geqslant 0} \sum^n_{m=0} (-1)^n \binom{n}{m} \partial_q^{n-m}  h^*_n(q,t)  \partial_q^m \psi(q,t).
\label{a6}
\end{equation}
Using the identity $\sum_{n \geqslant 0} \sum^n_{m=0} f_{nm} =\sum_{n \geqslant 0} \sum_{m \geqslant n} f_{mn}$ we have
\begin{equation}
{\widehat H}(q,\partial_q,t)\psi(q,t) = \sum_{n \geqslant 0}\sum_{m \geqslant n}  (-1)^m \binom{m}{n} \partial_q^{m-n}  h^*_m(q,t)  \partial_q^n\psi(q,t) .
\label{a7}
\end{equation}
Comparing this expression with \eqref{2} gives 
\begin{equation}
h_n = \sum_{m \geqslant n} (-1)^m  \binom{m}{n} \partial_q^{m-n}  h^*_m.
\label{a8}
\end{equation}
Conversely, the condition \eqref{a8} implies that 
\begin{equation}
\sum_{n\geqslant 0} h_n(q,t) \partial_q^n \psi(q,t) =\sum_{n \geqslant 0} (-\partial_q)^n \left( h^*_n(q,t)\psi(q,t) \right)
\end{equation}
so that the Hamiltonian is Hermitian.
\Endproof

\bigskip

\noindent {\it Proof of Identity~\ref{iden1}.}
The identity is proved by working out the right-hand side of equation \eqref{6}:
\begin{equation}
\partial_q \left( \sum^{n-1}_{m=0} (-1)^{m} \partial^m_q \phi \partial^{n-m-1}_q \chi \right)= \sum^{n-1}_{m=0} (-1)^{m} \left( \partial^{m+1}_q \phi \partial^{n-m-1}_q \chi + \partial^m_q \phi \partial^{n-m}_q \chi \right).
\label{a1}
\end{equation}
Using the identities
\begin{equation}
\sum^{n-1}_{m=0} (-1)^{m} \partial^{m+1}_q \phi \partial^{n-m-1}_q \chi = \sum^{n-2}_{m=0} (-1)^{m} \partial^{m+1}_q \phi \partial^{n-m-1}_q \chi - (-1)^{n}\chi \partial^n_q \phi
\label{a2}
\end{equation}
and
\begin{align}
\sum^{n-1}_{m=0} (-1)^{m} \partial^m_q \phi \partial^{n-m}_q \chi &= \sum^{n-2}_{m=-1} (-1)^{m+1} \partial^{m+1}_q \phi \partial^{n-m-1}_q \chi \nonumber\\
 &= - \sum^{n-2}_{m=0} (-1)^{m} \partial^{m+1}_q \phi \partial^{n-m-1}_q \chi  + \phi \partial^n_q \chi ,
\label{a3}
\end{align}
we find that \eqref{a1} becomes $\phi \partial^n_q \chi - (-1)^{n}\chi \partial^n_q \phi$, which is the left-hand side of equation \eqref{6}.  
\Endproof

\bigskip

\noindent {\it Proof of equation \eqref{15}.}
Regarding $j^\psi$ as a function of the variables $\partial^n_q \psi \partial^m_q \psi^*$, we have that $J_{nm} = \partial j^\psi / \partial (\partial^n_q \psi \partial^m_q \psi^*)$.  To calculate $\partial j^\psi / \partial (\partial^n_q \psi \partial^m_q \psi^*)$, we first use Leibniz' rule to write 
\begin{align}
j^\psi  &= \ui\sum_{r \geqslant 1} \sum^{r-1}_{s=0}  (-1)^s \partial^s_q(\psi^*  h_r ) \partial^{r-s-1}_q \psi \nonumber\\
&=  \ui \sum_{r \geqslant 1} \sum^{r-1}_{s=0} \sum^{s}_{t=0} (-1)^s \binom{s}{t} \partial^t_q \psi^*  \partial^{s-t}_q h_r  \partial^{r-s-1}_q \psi.
\label{a9}
\end{align}
We then have
\begin{align}
\frac{\partial j^\psi }{ \partial (\partial^n_q \psi \partial^m_q \psi^*)} &= \ui \sum_{r \geqslant 1} \sum^{r-1}_{s=0} \sum^{s}_{t=0} (-1)^s \binom{s}{t}  \partial^{s-t}_q h_r   \delta_{n,r-s-1}\delta_{m,t} \nonumber\\
&= \ui \sum_{r \geqslant m+1} \sum^{r-1}_{s=m} (-1)^s \binom{s}{m} \partial^{s-m}_q h_r \delta_{n,r-s-1} \nonumber\\
&= \ui\sum_{r \geqslant n+m+1}  (-1)^{r+n+1} \binom{r-n-1}{m} \partial^{r-n-m-1}_q h_r .
\label{a10}
\end{align}
\Endproof

\bigskip

\noindent {\it Proof that the current $j^\psi$ given by \eqref{13} is real.}
The proof is best given by starting from the form \eqref{14} of the current. The current is real if and only if $J_{nm}=J^*_{mn}$ (for all $n,m \geqslant 0$). 

Before proving the latter, we first show that:
\begin{equation}
\sum^r_{s=n+m+1} (-1)^s \frac{(s-n-1)!}{s!} \binom{r -n-m-1}{r-s} = (-1)^{n+m+1} \frac{m!(r-m-1)!}{r!n!},  
\label{a12}
\end{equation}
for $r \geqslant n+m+1$, with $r,n,m \in {\mathbb N}$. We prove this by induction on $l=r-n-m-1 \geqslant 0$. For $l=0$, we have that both sides of equation \eqref{a12} are equal to $(-1)^r m!/r!$. Suppose now $l > 0$ and suppose the identity holds for all smaller values of $l$. Using the identity 
\begin{equation}
\binom{r -n-m-1}{r-s} = \binom{r -n-m-2}{r-s} +  \binom{r -n-m-2}{r-s-1},
\label{a13}
\end{equation}
which is valid for any $r,n,m$ and $s$ \cite[p.\ 174]{graham90},{\footnote{We use the definition of binomial coefficients in ref.\ \cite[p.\ 154]{graham90}, which reads 
\begin{equation}
\binom{n}{m} = \left\{ \begin{array}{ll}
\frac{n(n-1)\cdots(n-m+1)}{m(m-1)\cdots1}, & m > 0\\
1, & m = 0\\
0, & m < 0
\end{array} \right.  
\label{a13.1}
\end{equation}
for integer $n,m$.}} the sum on the left-hand side of \eqref{a12} decomposes into two sums, which we call $S_1$ and $S_2$. Let us first consider
\begin{equation}
S_1 = \sum^r_{s=n+m+1} (-1)^s \frac{(s-n-1)!}{s!} \binom{r -n-m-2}{r-s}.
\label{a14}
\end{equation}
Since $\binom{r -n-m-2}{r-s}=0$ for $s=n+m+1$ and $r-n-m-1 > 0$, we have
\begin{equation}
S_1 = \sum^r_{s=n+m+2} (-1)^s \frac{(s-n-1)!}{s!} \binom{r -n-m-2}{r-s}.
\label{a15}
\end{equation}
Applying the induction hypothesis we find
\begin{equation}
S_1 = (-1)^{n+m} \frac{(m+1)!(r-m-2)!}{r!n!}.
\label{a16}
\end{equation}
Similarly we have
\begin{align}
S_2 &=  \sum^r_{s=n+m+1} (-1)^s \frac{(s-n-1)!}{s!} \binom{r -n-m-2}{r-s-1} \nonumber\\
 &=  \sum^{r-1}_{s=n+m+1} (-1)^s \frac{(s-n-1)!}{s!} \binom{r -n-m-2}{r-s-1} \nonumber\\
&=  (-1)^{n+m+1} \frac{m!(r-m-2)!}{(r-1)!n!}.
\label{a17}
\end{align}
Combining these results we find that the left-hand side of equation \eqref{a12} equals
\begin{align}
S_1 + S_2 &= (-1)^{n+m+1} \frac{m!(r-m-1)!}{r!n!} \left( - \frac{m+1}{r-m-1} + \frac{r}{r-m-1}  \right)\nonumber\\
&= (-1)^{n+m+1} \frac{m!(r-m-1)!}{r!n!}, \nonumber\\
\label{a18}
\end{align}
which is equal to the right-hand side of \eqref{a12}. This completes the proof of the identity \eqref{a12}. 

Consider now
\begin{equation}
J_{nm} = \ui \sum_{r \geqslant n+m+1} (-1)^{r+n+1} \binom{r-n-1}{m} \partial^{r-n-m-1}_q h_r .
\label{a19}
\end{equation}
Using respectively the Hermiticity condition \eqref{3}, the identity $\sum_{r \geqslant n+m+1} \sum_{s \geqslant r}  f_{rs} = \sum_{r \geqslant n+m+1} \sum^r_{s =  n+m+1}  f_{sr}$, the factorial form of the binomials, and the identity \eqref{a12}, we find 
\begin{align}
J_{nm} &= \ui\sum_{r \geqslant n+m+1} \sum_{s \geqslant r}  (-1)^{r+n +s+1} \binom{r-n-1}{m} \binom{s}{r} \partial^{s-n-m-1}_q h^*_s \nonumber\\
&= \ui\sum_{r \geqslant n+m+1} \sum^r_{s =  n+m+1}   (-1)^{r+n+s+1} \binom{s-n-1}{m} \binom{r}{s} \partial^{r-n-m-1}_q h^*_r \nonumber\\
&= \ui\sum_{r \geqslant n+m+1} (-1)^{r+n+1} \frac{r!}{m!(r-n-m-1)!} \partial^{r-n-m-1}_q  h^*_r \nonumber\\
&\qquad \times \sum^r_{s=n+m+1} (-1)^s \frac{(s-n-1)!}{s!} \binom{r -n-m-1}{r-s} \nonumber\\
&= \ui\sum_{r \geqslant n+m+1} (-1)^{r+m} \binom{r-m-1}{n} \partial^{r-n-m-1}_q h^*_r \nonumber\\
&= J^*_{mn}.
\label{a20}
\end{align}
\Endproof

%%%%%%%%%%%%%%%%%%%%%%%%%%%%%%%%%%%%%%%%%%%%%%%%%%%%%%%%%%%%%%%%%%%%%%%%%%%%%%%%%%%%%%%%%%%%%%%%%%%%%%%%%%%%%%%%%%5
%%%%%%%%%%%%%%%%%%%%%%%%%%%%%%%%%%%%%%%%%%%%%%%%%%%%%%%%%%%%%%%%%%%%%%%%%%%%%%%%%%%%%%%%%%%%%%%%%%%%%%%%%%%%%%%%%%5
%%%%%%%%%%%%%%%%%%%%%%%%%%%%%%%%%%%%%%%%%%%%%%%%%%%%%%%%%%%%%%%%%%%%%%%%%%%%%%%%%%%%%%%%%%%%%%%%%%%%%%%%%%%%%%%%%%5

\section{Proofs for an $N$-dimensional configuration space}  \label{appmulti}
\noindent {\it Proof of the Hermiticity condition \eqref{53}.}
From \eqref{52.1}, and using Leibniz' rule, we have
\begin{equation}
{\widehat H}(q,D,t)\psi(q,t) = \sum_{n \geqslant 0} \sum_{0 \leqslant m \leqslant n} (-1)^{|n|} \binom{n}{m} D^{n-m}  h^*_n(q,t)  D^m\psi(q,t) .
\label{a51}
\end{equation}
Using the identity $\sum_{n \geqslant 0} \sum_{0 \leqslant m \leqslant n} f_{nm} =\sum_{n \geqslant 0} \sum_{m \geqslant n} f_{mn}$ we have
\begin{equation}
{\widehat H}(q,D,t)\psi(q,t) = \sum_{n \geqslant 0}\sum_{m \geqslant n}  (-1)^{|m|} \binom{m}{n} D^{m-n}  h^*_m(q,t) D^n\psi(q,t) .
\label{a52}
\end{equation}
Comparing this expression with \eqref{52} gives 
\begin{equation}
h_n = \sum_{m \geqslant n} (-1)^{|m|}  \binom{m}{n} D^{m-n}  h^*_m.
\label{a53}
\end{equation}
Conversely, the condition \eqref{a53} implies that 
\begin{equation}
\sum_{n \geqslant 0} h_n(q,t) D^n \psi(q,t)= \sum_{n \geqslant 0} (-1)^{|n|} D^n \left( h^*_n(q,t)\psi(q,t) \right),
\label{a54}
\end{equation}
so that the Hamiltonian is Hermitian.
\Endproof

\bigskip

\noindent {\it Proof of Identity~\ref{iden2}.}
The identity is proved by working out the right-hand side of \eqref{56}:
\begin{multline}
\sum^N_{i=1} D^{e_i} \left( \sum_{0 \leqslant m \leqslant n - e_i} (-1)^{|m|} \frac{n!}{|n|!} \frac{|m|!}{m!} \frac{|n-m-e_i|!}{(n-m-e_i)!}  D^m \phi D^{n - m - e_i} \chi \right) =\\
 \sum^N_{i=1}  \sum_{0 \leqslant m \leqslant n - e_i} (-1)^{|m|} \frac{n!}{|n|!}  \frac{|m|!}{m!} \frac{|n-m-e_i|!}{(n-m-e_i)!}  \left( D^{m+e_i} \phi D^{n - m - e_i} \chi + D^m \phi D^{n - m} \chi \right).
\label{a55}
\end{multline}
We can write
\begin{align}
&\sum_{0 \leqslant m \leqslant n - e_i} (-1)^{|m|}  \frac{|m|!}{m!} \frac{|n-m-e_i|!}{(n-m-e_i)!}  D^{m+e_i} \phi D^{n - m - e_i} \chi \nonumber\\
&\qquad = \sum_{e_i \leqslant m \leqslant n } (-1)^{|m-e_i|}  \frac{|m-e_i|!}{(m-e_i)!} \frac{|n - m|!}{(n - m)!}  D^{m} \phi D^{n - m} \chi \nonumber\\
&\qquad = \sum_{e_i \leqslant m \leqslant n } (-1)^{|m|+1}  \frac{|m|!}{m!}  \frac{|n - m|!}{(n - m)!}  \frac{m_i}{|m|} D^{m} \phi D^{n - m} \chi \nonumber\\
&\qquad = \sum_{0 < m \leqslant n } (-1)^{|m|+1}  \frac{|m|!}{m!}  \frac{|n - m|!}{(n - m)!}  \frac{m_i}{|m|} D^{m} \phi D^{n - m} \chi \nonumber\\
&\qquad = \sum_{0 < m < n } (-1)^{|m|+1}  \frac{|m|!}{m!}  \frac{|n - m|!}{(n - m)!}  \frac{m_i}{|m|} D^{m} \phi D^{n - m} \chi - (-1)^{|n|} \frac{|n|!}{n!}   \frac{n_i}{|n|} D^n \phi  \chi.
\label{a56}
\end{align}
(We can extend the range of the sum in the second-last line, because the terms with $m_i=0$ are zero.) Similarly, we can write
\begin{align}
&\sum_{0 \leqslant m \leqslant n - e_i} (-1)^{|m|}  \frac{|m|!}{m!} \frac{|n-m-e_i|!}{(n-m-e_i)!}  D^{m} \phi D^{n - m} \chi \nonumber\\
&\qquad = \sum_{0 \leqslant m \leqslant n - e_i} (-1)^{|m|}  \frac{|m|!}{m!} \frac{|n - m|!}{(n - m)!} \frac{(n_i-m_i)}{|n-m|} D^{m} \phi D^{n - m} \chi \nonumber\\
&\qquad = \sum_{0 \leqslant m < n } (-1)^{|m|}  \frac{|m|!}{m!} \frac{|n - m|!}{(n - m)!} \frac{(n_i-m_i)}{|n-m|} D^{m} \phi D^{n - m} \chi \nonumber\\
&\qquad = \sum_{0 < m < n } (-1)^{|m|}  \frac{|m|!}{m!} \frac{|n - m|!}{(n - m)!} \frac{(n_i-m_i)}{|n-m|} D^{m} \phi D^{n - m} \chi + \frac{|n|!}{n!}   \frac{n_i}{|n|} \phi  D^n \chi.
\label{a57}
\end{align}
Using \eqref{a55}, \eqref{a56} and \eqref{a57}, the right-hand side of \eqref{56} becomes
\begin{multline}
\phi D^n \chi - (-1)^{|n|}    \chi  D^n  \phi\\
+ \sum^N_{i=1} \sum_{0 < m < n } (-1)^{|m|} \frac{n!}{|n|!}  \frac{|m|!}{m!} \frac{|n - m|!}{(n - m)!} \left(  \frac{(n_i-m_i)}{|n-m|}  - \frac{m_i}{|m|} \right) D^{m} \phi D^{n - m} \chi .
\label{a58}
\end{multline}
Using  
\begin{equation}
\sum^N_{i=1} \left(  \frac{(n_i-m_i)}{|n-m|}  - \frac{m_i}{|m|} \right) = 0,
\label{a59}
\end{equation}
\eqref{a58} is equal to $\phi D^n \chi - (-1)^{|n|}    \chi  D^n  \phi$, which is just the left-hand side of \eqref{56}.
\Endproof

\bigskip

\noindent {\it Proof of equation \eqref{60}.}
Regarding the $j^\psi_i$ as functions of the variables $D^n \psi D^m \psi^*$, we have that $J_{i,nm} = \partial j^\psi_i / \partial (D^n \psi D^m \psi^*)$.  To calculate $\partial j^\psi_i / \partial (D^n \psi D^m \psi^*)$, we first use Leibniz' rule to write 
\begin{align}
j^\psi_i  &= \ui \sum_{r \geqslant e_i}  \sum_{0 \leqslant s \leqslant r - e_i} (-1)^{|s|} \frac{r!}{|r|!}  \frac{|s|!}{s!} \frac{|r-s-e_i|!}{(r-s-e_i)!}  D^s (\psi^* h_r)  D^{r - s -e_i} \psi  \nonumber\\
&= \ui \sum_{r \geqslant e_i} \sum_{0 \leqslant s \leqslant r - e_i} \sum_{0 \leqslant t \leqslant s}  (-1)^{|s| } \frac{r!}{|r|!}  \frac{|s|!}{s!} \frac{|r-s-e_i|!}{(r-s-e_i)!} \binom{s}{t}  D^t \psi^*  D^{s-t} h_r  D^{r-s-e_i} \psi.
\label{a61}
\end{align}
We then have
\begin{align}
\frac{\partial j^\psi_i}{\partial (D^n \psi D^m \psi^*)} &=  \ui \sum_{r \geqslant e_i} \sum_{0 \leqslant s \leqslant r - e_i} \sum_{0 \leqslant t \leqslant s}  (-1)^{|s| } \frac{r!}{|r|!}  \frac{|s|!}{s!} \frac{|r-s-e_i|!}{(r-s-e_i)!} \binom{s}{t}   D^{s-t} h_r \delta_{n,r-s-e_i} \delta_{m,t}\nonumber\\
&= \ui \sum_{r \geqslant m+e_i} \sum_{m \leqslant s \leqslant r - e_i}  (-1)^{|s|} \frac{r!}{|r|!}  \frac{|s|!}{s!} \frac{|r-s-e_i|!}{(r-s-e_i)!} \binom{s}{m}  D^{s-m} h_r   \delta_{n,r-s-e_i}\nonumber\\
&= \ui \sum_{r \geqslant n+m+e_i}  (-1)^{|r+n|+1}  \frac{r!}{|r|!} \frac{|r-n-e_i|!}{(r-n-e_i)!} \frac{|n|!}{n!} \binom{r-n-e_i}{m} D^{r-n-m-e_i} h_r .
\label{a62}
\end{align}
\Endproof

\bigskip

\noindent {\it Proof that the currents $j^\psi_i$ given by \eqref{58} are real.}
The proof is best given by starting from the form \eqref{59} of the currents. The currents are real if and only if $J_{i,nm}=J^*_{i,mn}$ (for all $n,m \geqslant 0$). 

Before proving the latter, we first show that:
\begin{equation}
\sum_{n+m+e_i \leqslant s \leqslant r} (-1)^{|s|} \frac{|s-n-e_i|!}{|s|!} \binom{r -n-m- e_i}{r-s} = (-1)^{|n+m|+1} \frac{|m|!|r-m-e_i|!}{|r|!|n|!},  
\label{a90}
\end{equation}
for $r \geqslant n+m+e_i$, with $r,n,m \in {\mathbb N}^N_0$. We prove this by induction on $l=r-n-m-e_i \geqslant 0$. For $l=0$, we have that both sides of equation \eqref{a90} are equal to $(-1)^{|r|} |m|!/|r|!$. Suppose now $l > 0$ and suppose the identity holds for all smaller values of $l$. Since $l > 0$ there exists a $j\in \{1,\dots,N\}$ such that $l - e_j = r-n-m-e_i -e_j \geqslant 0$. Using the identity 
\begin{equation}
\binom{r -n-m-e_i}{r-s} = \binom{r -n-m-e_i-e_j}{r-s} +  \binom{r -n-m-e_i-e_j}{r-s-e_j},
\label{a91}
\end{equation}
which is valid for any $r,n,m$ and $s$ \cite[p.\ 174]{graham90}, the sum on the left-hand side of \eqref{a90} decomposes into two sums, which we call $S_1$ and $S_2$. Let us first consider
\begin{equation}
S_1 = \sum_{n+m+e_i \leqslant s \leqslant r} (-1)^{|s|} \frac{|s-n-e_i|!}{|s|!} \binom{r -n-m-e_i-e_j}{r-s}.
\label{a92}
\end{equation}
Since $\binom{r -n-m-e_i-e_j}{r-s}=0$ for $s_j=n_j+m_j+\delta_{ij}$ and $r-n-m-e_i-e_j \geqslant 0$, we have
\begin{equation}
S_1 = \sum_{n+m+e_i+e_j \leqslant s \leqslant r} (-1)^{|s|} \frac{|s-n-e_i|!}{|s|!} \binom{r -n-m-e_i-e_j}{r-s}.
\label{a93}
\end{equation}
Applying the induction hypothesis we find
\begin{equation}
S_1 = (-1)^{|n+m|} \frac{|m+e_j|!|r-m-e_i-e_j|!}{|r|!|n|!}.
\label{a94}
\end{equation}
Similarly we have
\begin{align}
S_2 &=  \sum_{n+m+e_i \leqslant s \leqslant r} (-1)^{|s|} \frac{|s-n-e_i|!}{|s|!} \binom{r -n-m-e_i-e_j}{r-s-e_j} \nonumber\\
 &=  \sum_{n+m+e_i \leqslant s \leqslant r-e_j} (-1)^{|s|} \frac{|s-n-e_i|!}{|s|!} \binom{r -n-m-e_i-e_j}{r-s-e_j} \nonumber\\
&=  (-1)^{|n+m|+1} \frac{|m|!|r-m-e_i-e_j|!}{|r-e_j|!|n|!}.
\label{a95}
\end{align}
Combining these results we find that the left-hand side of equation \eqref{a90} equals
\begin{align}
 S_1 + S_2 &= (-1)^{|n+m|+1} \frac{|m|!|r-m-e_i|!}{|r|!|n|!} \left( - \frac{|m+e_j|}{|r-m-e_i|} + \frac{|r|}{|r-m-e_i|}  \right)\nonumber\\
&= (-1)^{|n+m|+1} \frac{|m|!|r-m-e_i|!}{|r|!|n|!},
\label{a96}
\end{align}
which is equal to the right-hand side of \eqref{a90}. This completes the proof of the identity \eqref{a90}. 

Consider now
\begin{equation}
J_{i,nm} =\ui\sum_{r \geqslant n+m+e_i} (-1)^{|r+n|+1}  \frac{r!}{|r|!} \frac{|r-n-e_i|!}{(r-n-e_i)!} \frac{|n|!}{n!} \binom{r-n-e_i}{m} D^{r-n-m-e_i} h_r 
\label{a97}
\end{equation}
Using respectively the Hermiticity condition \eqref{53}, the identity $\sum_{r \geqslant n+m+e_i} \sum_{s \geqslant r}  f_{rs} = \sum_{r \geqslant n+m+e_i} \sum_{n+m+e_i \leqslant s \leqslant r}  f_{sr}$, the factorial form of the binomials, and the identity \eqref{a90}, we find 
\begin{align}
J_{i,nm} &= \ui \sum_{r \geqslant n+m+e_i} \sum_{s \geqslant r}  (-1)^{|r+n+s|+1}  \frac{r!}{|r|!} \frac{|r-n-e_i|!}{(r-n-e_i)!} \frac{|n|!}{n!} \binom{r-n-e_i}{m} \binom{s}{r} D^{s-n-m-e_i} h^*_s \nonumber\\
&= \ui \sum_{r \geqslant n+m+e_i}  \sum_{n+m+e_i \leqslant s \leqslant r}   (-1)^{|r+n+s|+1} \nonumber\\
&\qquad \times \frac{s!}{|s|!} \frac{|s-n-e_i|!}{(s-n-e_i)!} \frac{|n|!}{n!}  \binom{s-n-e_i}{m} \binom{r}{s} D^{r-n-m-e_i} h^*_r \nonumber\\
&= \ui\sum_{r \geqslant n+m+e_i}  (-1)^{|r+n|+1} \frac{|n|!r!}{n!m!(r-n-m-e_i)!} D^{r-n-m-e_i}  h^*_r \nonumber\\
&\qquad \times \sum_{n+m+e_i \leqslant s \leqslant r} (-1)^{|s|} \frac{|s-n-e_i|!}{|s|!} \binom{r -n-m-e_i}{r-s} \nonumber\\
&= \ui \sum_{r \geqslant n+m+e_i}  (-1)^{|r+m|} \frac{r!}{|r|!} \frac{|r-m-e_i|!}{(r-m-e_i)!} \frac{|m|!}{m!} \binom{r-m-e_i}{n} D^{r-n-m-e_i} h^*_r \nonumber\\
&= J^*_{i,mn}.
\label{a98}
\end{align}
\Endproof

\end{document}